\documentclass[final,3p]{elsarticle}

\usepackage{amsmath}
\usepackage{amssymb}

\journal{Annals of Physics}

\begin{document}

\begin{frontmatter}


\title{Shock Wave Polarizations and Optical Metrics in the Born and the Born-Infeld Electrodynamics}
\author[label1]{Christoph Minz}
\ead{christoph.minz@alumni.tu-berlin.de}
\author[label1]{Horst-Heino von Borzeszkowski}
\ead{borzeszk@mailbox.tu-berlin.de}
\author[label1]{Thoralf Chrobok}
\ead{tchrobok@mailbox.tu-berlin.de}
\author[label2]{Gerold Schellstede}
\ead{schellst@physik.fu-berlin.de}
\address[label1]{Institute of Theoretical Physics, Technische Universit\"at Berlin, Hardenbergstr. 36, D-10623 Berlin, Germany}
\address[label2]{ZARM (Center of Applied Space Technology and Microgravity), Universit\"at Bremen, Am Fallturm, D-28359 Bremen, Germany}

\begin{abstract}
We analyze the behavior of shock waves in nonlinear theories of electrodynamics. 
For this, by use of generalized Hadamard step functions of increasing order, the electromagnetic potential is developed in a series expansion near the shock wave front. 
This brings about a corresponding expansion of the respective electromagnetic field equations which allows for deriving relations that determine the jump coefficients in the expansion series of the potential. 
We compute the components of a suitable gauge-normalized version of the jump coefficients given for a prescribed tetrad compatible with the shock front foliation. 
The solution of the first-order jump relations shows that, in contrast to linear Maxwell's electrodynamics, in general the propagation of shock waves in nonlinear theories is governed by optical metrics and polarization conditions describing the propagation of two differently polarized waves (leading to a possible appearance of birefringence). 
In detail, shock waves are analyzed in the Born and Born-Infeld theories verifying that the Born-Infeld model exhibits no birefringence and the Born model does.
The obtained results are compared to those ones found in literature. 
New results for the polarization of the two different waves are derived for Born-type electrodynamics. 
\end{abstract}

\begin{keyword}

\PACS 03.50.Kk \sep 41.20.Jb \sep 42.15.Dp

\end{keyword}

\end{frontmatter}

\section{Introduction}
\label{Introduction}
Foundation and study of nonlinear electrodynamic theories go back to ideas of Gustav Mie. 
The first gauge-invariant version of nonlinear electromagnetic field equations was presented by Born~\cite{1933born}; soon after, his ansatz was generalized by him and Infeld~\cite{1933borninfeld,1934borninfeld}. 
As an alternative to the linear field equations of Maxwell's electrodynamics, the new equations should provide solutions that can be interpreted as classical models of electrons. 
Another type of nonlinear field equation was introduced by Heisenberg and Euler~\cite{1936heisenbergeuler}. 
They showed that radiative quantum-electrodynamic effects semi-classically could be described by nonlinear correction terms to the linear classical equations. 
Nowadays, nonlinear electrodynamics also attract the interest because Born-Infeld-like actions arise as effective actions in superstring theory~\cite{1987bergshoeffetal,1987metsaevetal}.

Whatever the reason for the interest in nonlinear electrodynamics was, again and again there appeared papers on different aspects of such theories. 
In particular, this concerns the analysis of the propagation of waves. 
This goes back to the early 1950s and wins new attention not only of theorists, but - due to improved experimental technologies - also of experimental physicists. 
Early theoretical analysis was published by Boillat who found that both polarization modes travel along the light cone of one optical metric in exceptional nonlinear electrodynamics like in the Born-Infeld theory~\cite{1970boillat,1972boillat}. 
In this context, \emph{exceptional} means that the fields on the wave-front always satisfy the field equations. 
In a later paper Boillat and Ruggeri~\cite{2004boillatruggeri} extended the approach to the motion of more general discontinuity fronts. 
Considering the energy-momentum tensor under the assumption of convexity of energy, they derived relations concerning exceptional waves and shock fronts, where in the latter case there is a discontinuity of the field itself. 
Particularly, it was shown that characteristic shocks (i.e., shocks moving with the wave velocity) are unbounded (i.e., allowing arbitrarily large coefficients) for the Born-Infeld electrodynamics. 

Motivated by the theory of Hehl and Obukhov~\cite{2003hehlobukhov} according to which the constitutive law connecting the electromagnetic field strength with the electromagnetic excitation plays the role of a space-time relation defining the metric, Perlick chose a quite different approach~\cite{2011perlick}. 
He started with the metric-free formulation of Maxwell's equations and asked which constitutive laws yield a well-posed initial-value problem. 
Especially, he showed that the Born-Infeld theory admits a well-posed initial-value problem and has no birefringence which agrees with our results, although obtained with a different method. 

There are mainly two methods used in investigations of nonlinear electrodynamics, the method of geometric optics and, alternatively, the shock wave method which is based on Hadamard's theory of discontinuities~\cite{1903hadamard}. 
General aspects of the shock wave method can be found in Courant and Hilbert~\cite{1937couranthilbert} (see also the English translation~\cite{1962couranthilbert}) as well as Truesdell and Toupin~\cite{1960truesdelltoupin}. 
In the following, we shall use a modified shock wave method based on Treder~\cite{1962treder}. 
It (i) starts from discontinuities in the electromagnetic potential, instead from those ones of the field strengths, and (ii) develops the field quantities in a series expansion in the neighborhood of the jump hypersurface, where, following Stellmacher~\cite{1938stellmacher}, the jump relations of the different approximation steps are solved by contracting them with a suitably chosen tetrad field. 

The paper is organized as follows\footnote{Details especially on the mathematical calculations can be found in Ref.~\cite{2014minz}.}. 
After providing fundamentals of nonlinear electrodynamics, describing the used shock wave method in brief, and introducing the tetrad field, the electromagnetic potential is developed in a series expansion near the jump hypersurface which has to satisfy the field equations in all orders; this means that we consider exceptional waves in the sense of Boillat~\cite{1970boillat,1972boillat}.   
Then, by inserting this series in the field equations and performing the appropriate limits procedure, the jump relations can be calculated up to an arbitrary order (see Ref.~\cite{2014minz}). 
The first-order relations are solved by tetrad contraction. 
The solutions reproduce results of Obukhov and Rubilar~\cite{2002obukhovrubilar} (also discussed by Novello et al.~\cite{2000novelloetal}) concerning the so-called optical metric (sometimes referred to as effective metric). 
Furthermore, as a new result we calculate the polarization angle of the first-order jump relations and the possible appearance of birefringence for Born-type electrodynamics. 

We make use of the index notation with small greek letters running from 0 to 3. 
Indices are contracted with the Minkowski metric 
\begin{equation}
			\left( \eta_{\alpha \beta} \right) 
	= \textup{diag}(1, -1, -1, -1). 
	\label{eq:signature}
\end{equation}
The electromagnetic field tensor is defined as the exterior derivative of the electromagnetic potential, 
\begin{equation}
			F_{\alpha \beta} 
	= F_{[\alpha \beta]} 
	:= A_{\beta,\alpha} 
		- A_{\alpha,\beta}. 
	\label{eq:Field}
\end{equation}
With the fully antisymmetric Levi-Civita tensor~$\epsilon^{\alpha \beta \gamma \delta}$ ($\epsilon^{0123} = 1$), the dual field tensor reads 
\begin{equation}
			\overset{*}{F}{}^{\alpha \beta} 
	:= \frac{1}{2} \epsilon^{\alpha \beta \gamma \delta} F_{\gamma \delta}. 
	\label{eq:DualField}
\end{equation}
All quantities are in the International System of Units, which is denoted by $[\dots]_{\textup{SI}}$, for example $[F^{\alpha \beta}]_{\textup{SI}} = [\overset{*}{F}{}^{\alpha \beta}]_{\textup{SI}} = \textup{T}$.
The units of the electromagnetic fields are converted to the cgs system as follows~\cite{1999jackson} 
\begin{equation}
			\left[ F^{\alpha \beta} \right]_{\textup{SI}} 
	= \left[ \sqrt{\frac{\mu_0}{4 \pi}} F^{\alpha \beta} \right]_{\textup{cgs}}. 
	\label{eq:SIcgs}
\end{equation}

\section{The Born and the Born-Infeld Electrodynamics}
\subsection{Lagrange Densities}
Lagrange densities of nonlinear electrodynamics are functions of the two field invariants 
\begin{subequations}
	\label{eq:FandG}
	\begin{eqnarray}
				F 
		&:=& \frac{1}{2} F_{\alpha \beta} F^{\alpha \beta}, 
		\label{eq:F}
	\\
				G 
		&:=& \frac{1}{4} F_{\alpha \beta} \overset{*}{F}{}^{\alpha \beta}. 
		\label{eq:G}
	\end{eqnarray}
\end{subequations}

In 1933, Born started the discussion on a new electromagnetic field theory~\cite{1933born} for which the Lagrange density can be written in the form 
\begin{equation}
			\mathcal{L}_{\textup{B}} 
	= - \frac{b^2}{\mu_0} 
			\left( 
				\sqrt{1 + \frac{F}{b^2}} 
			- 1 
			\right) 
	\label{eq:BLagrange}
\end{equation}
introducing $b$ as a new natural constant, $[b]_{\textup{SI}} = [F_{\alpha \beta}]_{\textup{SI}}$, and with the magnetic permeability $\mu_0$. 
This theory is known as the Born electrodynamics and was investigated in more detail by Born and Infeld~\cite{1933borninfeld}. 
One year later, Born and Infeld published the foundations of the new field theory~\cite{1934borninfeld}, the Born-Infeld electrodynamics, where they derived another Lagrange density for a free field, 
\begin{equation}
			\mathcal{L}_{\textup{BI}} 
	= - \frac{b^2}{\mu_0} 
			\left( 
				\sqrt{1 + \frac{F}{b^2} - \frac{G^2}{b^4}} 
			- 1 
			\right). 
	\label{eq:BILagrange}
\end{equation}
With any Lagrange density $\mathcal{L}$ ($[\mathcal{L}]_{\textup{SI}} = \textup{J} \textup{m}^{-3}$), the action integral $S$ ($[S]_{\textup{SI}} = \textup{J} \textup{s}$) over the space volume $V_3$ and time span $\Delta t$ reads 
\begin{equation}
			S 
	= \int_{\Delta t}{\int_{V_3}{\mathcal{L}\; \textup{d}^3 x}\; \textup{d} t}. 
	\label{eq:Action}
\end{equation}

\subsection{Field Momenta}
Using the Euler-Lagrange formalism with the Lagrange densities~(\ref{eq:BILagrange}) and (\ref{eq:BLagrange}), one finds the vacuum field equations~$D^{\alpha \beta}{}_{,\alpha} = 0$ with the displacement field tensor (electromagnetic field momentum) 
\begin{equation}
			D_{\textup{BI}}{}^{\alpha \beta} 
	:= \frac{\partial \mathcal{L}_{\textup{BI}}}{\partial F_{\gamma \delta}} 
		\frac{\partial F_{\gamma \delta}}{\partial A_{\alpha,\beta}} 
	= \frac{1}{\mu_0} 
			\frac{ 
				F^{\alpha \beta} 
			- \frac{G}{b^2} \overset{*}{F}{}^{\alpha \beta} 
			}{\sqrt{1 + \frac{F}{b^2} - \frac{G^2}{b^4}}} 
	\label{eq:BIFieldMomentum}
\end{equation}
and 
\begin{equation}
			D_{\textup{B}}{}^{\alpha \beta} 
	:= \frac{\partial \mathcal{L}_{\textup{B}}}{\partial F_{\gamma \delta}} 
		\frac{\partial F_{\gamma \delta}}{\partial A_{\alpha,\beta}} 
	= \frac{1}{\mu_0} 
			\frac{ 
				F^{\alpha \beta} 
			}{\sqrt{1 + \frac{F}{b^2}}}, 
	\label{eq:BFieldMomentum}
\end{equation}
respectively. 

In case of the Born and the Born-Infeld electrodynamics, the field equations can be rewritten so that the square root disappears.

\subsection{Rewritten Field Equations}
In order to avoid a Taylor expansion of the square root in the shock wave formalism and to compare it to the linear field equations of Maxwell, they are multiplied with the third power of the square root. 
This mathematical reformulation does not change the results for the first-order but especially simplifies higher-order jump relations. 

For the Born-Infeld field equations, one defines 
\begin{equation}
			\phi_{\textup{BI}}{}^{\beta} 
	:= \mu_0 \left( 
			1 
		+ \frac{F}{b^2} 
		- \frac{G^2}{b^4} \right)^{\frac{3}{2}} 
			\underbrace{D_{\textup{BI}}{}^{\alpha \beta}{}_{,\alpha}}_{= 0}, 
	\label{eq:BIFieldEquationsSimplification}
\end{equation}
and, analogously, for Born's field equations 
\begin{equation}
			\phi_{\textup{B}}{}^{\beta} 
	:= \mu_0 \left( 
			1 
		+ \frac{F}{b^2} \right)^{\frac{3}{2}} 
			\underbrace{D_{\textup{B}}{}^{\alpha \beta}{}_{,\alpha}}_{= 0}. 
	\label{eq:BFieldEquationsSimplification}
\end{equation}

In more detail, the vanishing vector field~$\phi^{\beta}$ (with index BI for the Born-Infeld theory and index B for the Born electrodynamics) reads 
\begin{eqnarray}
			\phi_{\textup{BI}}{}^{\beta} 
	&=& F^{\alpha \beta}{}_{,\alpha} 
		+ \frac{1}{b^2} 
			\left( \vphantom{\frac{1}{2}} 
				F F^{\alpha \beta}{}_{,\alpha} 
			- \frac{1}{2} F^{\alpha \beta} F_{,\alpha} 
			- \overset{*}{F}{}^{\alpha \beta} G_{,\alpha} 
			\right) + \nonumber \\ && {}
		+ \frac{1}{b^4} 
			\left( 
			- G^2 F^{\alpha \beta}{}_{,\alpha} 
			+ \frac{1}{2} G \overset{*}{F}{}^{\alpha \beta} F_{,\alpha} 
			+ G F^{\alpha \beta} G_{,\alpha} 
			- F \overset{*}{F}{}^{\alpha \beta} G_{,\alpha} 
			\right) \qquad 
	\label{eq:BIFieldEquations}
\end{eqnarray}
and 
\begin{eqnarray}
			\phi_{\textup{B}}{}^{\beta} 
	&=& F^{\alpha \beta}{}_{,\alpha} 
		+ \underline{\frac{1}{b^2} F F^{\alpha \beta}{}_{,\alpha}} 
		- \underline{\underline{\frac{1}{2 b^2} F^{\alpha \beta} F_{,\alpha}}}, \qquad
	\label{eq:BFieldEquations}
\end{eqnarray}
respectively. 
For the calculations, the homogeneous field equations~$\overset{*}{F}{}^{\alpha \beta}{}_{,\alpha} = 0$ have been used which are also fulfilled for the Born and the Born-Infeld electrodynamics. 
The homogeneous field equations are mathematical identities when developed in a jump series~\cite{2014minz}. 

The first summand in equations~(\ref{eq:BIFieldEquations}) and (\ref{eq:BFieldEquations}) is the linear term known from Maxwell's electrodynamics. 
The nonlinear terms in Eq.~(\ref{eq:BFieldEquations}) are underlined to identify them in the jump series expansion~(\ref{eq:FieldXB}) below. 

To derive the full jump series expansion of the nonlinear electrodynamics, the shock wave formalism by Treder and Stellmacher will be used. 

\section{Shock Wave Formalism}
\subsection{Series Expansion of the Electromagnetic Potential}
The mathematical formulation of shock wave perturbations is based on the parameter~$\varSigma$ which specifies a position relative to the shock front in length units. 
The Eq.~$\varSigma = 0$ defines the shock front as a hypersurface in space-time. 
The electromagnetic potential~$A_{\alpha}$ can be expanded with respect to the parameter~$\varSigma$. 
According to Treder~\cite{1962treder}, this reads 
\begin{equation}
			A_{\alpha} 
	= A^-{}_{\alpha} 
		+ \sum_{m = l}^{\infty}{ \underset{\vphantom{l} m}{\varphi}{}_{\alpha} \underset{\vphantom{l} m}{\vphantom{f} h} }, 
	\label{eq:AX}
\end{equation}
where $A^-{}_{\alpha}\left( x^{\mu} \right)$ denotes the undisturbed background field which is a solution of the field equations. 
In the following, all background fields are marked by a minus symbol. 

The letter~$l$ always stands for the first index in the perturbation summation so that the first discontinuities occur for the~$l$-th derivative of the electromagnetic potential. 
However, $l$ has to be greater than~$1$, because for~$l = 1$ surface charges along~$\varSigma = 0$ have to be discussed~\cite{1961papapetroutreder}. 

\begin{figure}
	\centering
	\includegraphics{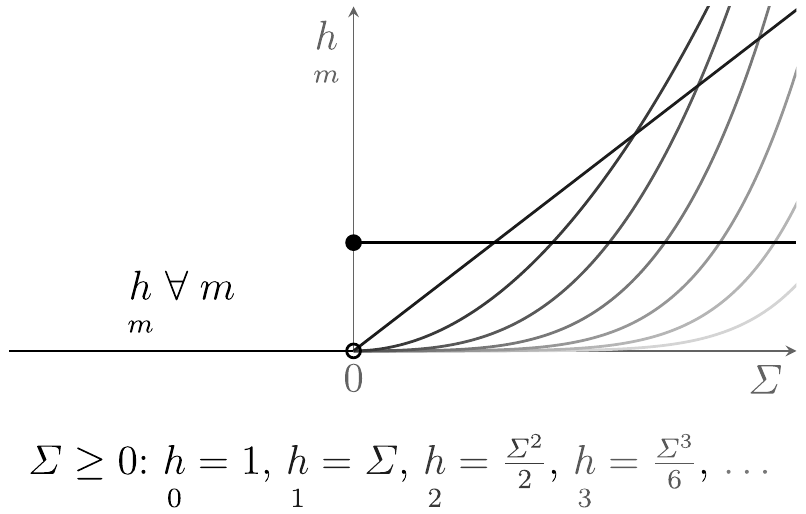}
	\caption{\label{fig:GeneralStepFunctions}General step functions.}
\end{figure}

The expansion~(\ref{eq:AX}) is similar to a conventional Taylor series. 
The coefficients~$\underset{\vphantom{l} m}{\varphi}{}_{\alpha} \left( x^{\mu} \right)$ are assumed to be analytic, since they do not depend on the expansion parameter~$\varSigma$, whereas~$\underset{\vphantom{l} m}{\vphantom{f} h} \left( \varSigma \right)$ represent general step functions discontinuous at~$\varSigma = 0$ in their~$m$-th derivative, 
\begin{equation}
			\underset{\vphantom{l} m}{\vphantom{f} h} \left( \varSigma \right) 
	:= \begin{cases}
			0 
			& \textup{ if } \varSigma < 0, \\
			\displaystyle \frac{\varSigma^m}{m!} 
			& \textup{ if } \varSigma \geq 0 
		\end{cases} 
	\label{eq:GeneralStepFunction}
\end{equation}
(see also FIG.~\ref{fig:GeneralStepFunctions}). 
When truncating the expression~(\ref{eq:AX}) at a certain number of summands, the field perturbation can be approximated in a small region~$\varSigma > 0$. 

The jump coefficients of the electromagnetic potential are 
\begin{equation}
			\underset{\vphantom{l} m}{\varphi}{}_{\alpha} 
	= \left[ \frac{\partial^m A_{\alpha}}{{\partial{\varSigma}}^m} \right]_{\varSigma = 0}, 
	\label{eq:AXCoefDef}
\end{equation}
where the derivatives with respect to the parameter $\varSigma$ are formal only. 
The jump brackets are defined as 
\begin{equation}
			\left[ J \right]_{\varSigma = 0} 
	:= \lim_{\varSigma \to +0}{\left( J \right)} 
		- \lim_{\varSigma \to -0}{\left( J \right)}. 
	\label{eq:JumpBrackets}
\end{equation}
$J$ stands for an arbitrary function~\cite{2000novelloetal}. 

\subsection{Series Expansion of the Electromagnetic Field}
Similarly to the potential series expansion~(\ref{eq:AX}), the field series expansion is given by 
\begin{equation}
			F_{\alpha \beta} 
	= F^-{}_{\alpha \beta} 
		+ \sum_{m = l'}^{\infty} 
			\underset{\vphantom{l} m}{f}{}_{\alpha \beta} 
			\underset{\vphantom{l} m}{\vphantom{f} h}, 
	\label{eq:FieldX}
\end{equation}
where the coefficients~$\underset{\vphantom{l} m}{f}{}_{\alpha \beta}$ are functions of~$\underset{\vphantom{l} m}{\varphi}{}_{\alpha}$ linking the Eq.~(\ref{eq:FieldX}) to the derivatives of~$A_{\alpha}$, 
\begin{subequations}
	\label{eq:FieldXCoefs}
	\begin{eqnarray}
				\underset{l - 1}{f}{}_{\alpha \beta} 
		&=& p_{\alpha} \underset{l}{\vphantom{f} \varphi}{}_{\beta} 
			- p_{\beta} \underset{l}{\vphantom{f} \varphi}{}_{\alpha}, 
		\label{eq:FieldXCoef1}
	\\
				\underset{\vphantom{l} m}{f}{}_{\alpha \beta} 
		&=& \underset{\vphantom{l} m}{\vphantom{f} \varphi}{}_{\beta,\alpha} 
			- \underset{\vphantom{l} m}{\vphantom{f} \varphi}{}_{\alpha,\beta} 
			+ p_{\alpha} \underset{\vphantom{l} m + 1}{\vphantom{f} \varphi}{}_{\beta} 
			- p_{\beta} \underset{\vphantom{l} m + 1}{\vphantom{f} \varphi}{}_{\alpha} 
			\quad (\forall\; m \geq l). 
		\label{eq:FieldXCoefm} 
	\end{eqnarray}
\end{subequations}
The new start index is~$l' = l - 1$ and the normal vector field 
\begin{equation}
			p_{\alpha} 
	:= \partial_{\alpha} \varSigma 
	\label{eq:TetradNormal}
\end{equation}
appears in the coefficients~(\ref{eq:FieldXCoefs}) due to the chain rule, 
\begin{equation}
			\partial_{\alpha} \underset{\vphantom{l} m}{\vphantom{f} h} 
	= p_{\alpha} \underset{\vphantom{l} m - 1}{\vphantom{f} h}. 
	\label{eq:GeneralStepFunctionNormalDeriv}
\end{equation}
$p_{\alpha}$ is perpendicular to the shock front~$\varSigma = 0$ in every point. 

\subsection{Tetrad Fields}
The tetrad which is used to formulate shock wave solutions and to contract the jump relations consists of four vector fields, the normal vector field~$p_{\alpha}$, the complex spacelike vector field~$\omega_{\alpha} = \textup{Re}(\omega_{\alpha}) + \textup{i} \textup{Im}(\omega_{\alpha})$, its conjugate~$\bar{\omega}_{\alpha}$, and a timelike vector field~$d_{\alpha}$. 
$\omega_{\alpha}$ was chosen complex to interpret the fields perpendicular to the propagation direction similar to optics with an amplitude and polarization angle. 

In contrast to the tetrad fields introduced by Stellmacher~\cite{1938stellmacher} and Treder~\cite{1962treder}, the scalar 
\begin{equation}
			p 
	:= p_{\alpha} p^{\alpha} 
	= \eta^{\alpha \beta} p_{\alpha} p_{\beta} 
	\label{eq:TetradNormalScalar}
\end{equation}
does not have to be zero. 
The four tetrad fields fulfill the conditions 
\begin{subequations}
	\label{eq:Tetrad}
	\begin{eqnarray}
				p_{\alpha} \omega^{\alpha} 
		= p_{\alpha} \bar{\omega}^{\alpha} 
		:= 0, \quad 
		&&		p_{\alpha} d^{\alpha} 
		:= d, \qquad
		\\
				\omega_{\alpha} \omega^{\alpha} 
		= \bar{\omega}_{\alpha} \bar{\omega}^{\alpha} 
		:= 0, \quad 
		&&		\omega_{\alpha} \bar{\omega}^{\alpha} 
		:= - \frac{1}{2}, \qquad
		\\ 
				\omega_{\alpha} d^{\alpha} 
		= \bar{\omega}_{\alpha} d^{\alpha} 
		:= 0, \quad 
		&&		d_{\alpha} d^{\alpha} 
		:= 1. \qquad
	\end{eqnarray}
\end{subequations}

For a better physical interpretation of the parts of the electromagnetic fields pointing in~$d_{\alpha}$-direction, one may choose~$d_{\alpha}$ so that~$p_{\alpha} d^{\alpha} = 0$ when~$p_{\alpha}$ is not lightlike. 
However, to avoid the discussion of two different cases~$p \neq 0$ (wave front traveling slower than the speed of light~$c$) and~$p = 0$ (wave front traveling with the maximum speed~$c$), relations~(\ref{eq:Tetrad}) will be used according to Treder's tetrad~\cite{1962treder}. 
In general $p_{\alpha}$ is timelike and specifies the value of $d$. 

For one step of the calculations, it is useful to write the Minkowskian metric in terms of the tetrad fields, 
\begin{equation}
			\eta_{\alpha \beta} 
	= \frac{p}{p - d^2} d_{\alpha} d_{\beta} 
		- \frac{1}{p - d^2} 
			\left( 
				d d_{\alpha} p_{\beta} 
			+ d p_{\alpha} d_{\beta} 
			- p_{\alpha} p_{\beta} 
			\right) 
		- 2 
			\left( 
				\omega_{\alpha} \bar{\omega}_{\beta} 
			+ \bar{\omega}_{\alpha} \omega_{\beta} 
			\right). 
	\label{eq:TetradsMetric}
\end{equation}

With the parameters shock excitation~$\underset{\vphantom{l} m}{\vphantom{f} \tilde{J}}$, polarization angle~$\underset{\vphantom{l} m}{\vphantom{f} \tilde{\kappa}}$, temporal amplitude~$\underset{\vphantom{l} m}{\vphantom{f} \tilde{\lambda}}$, and normal amplitude~$\underset{\vphantom{l} m}{\vphantom{f} a}$ the general solution of the~$m$-th-order coefficient ($m \geq l$) is a superposition of the four vector fields, 
\begin{equation}
			\underset{\vphantom{l} m}{\varphi}{}_{\alpha} 
	= \sqrt{\underset{\vphantom{l} m}{\vphantom{f} \tilde{J}}} 
			\left( 
				\textup{e}^{\textup{i} \underset{\vphantom{l} m}{\vphantom{f} \tilde{\kappa}}} \omega_{\alpha} 
			+ \textup{e}^{-\textup{i} \underset{\vphantom{l} m}{\vphantom{f} \tilde{\kappa}}} \bar{\omega}_{\alpha} 
			\right) 
		+ \underset{\vphantom{l} m}{\vphantom{f} \tilde{\lambda}} d_{\alpha} 
		+ \underset{\vphantom{l} m}{\vphantom{f} a} p_{\alpha}. 
	\label{eq:AXCoefGeneralSolution}
\end{equation}
By applying a discontinuous gauge transformation along $\varSigma = 0$, the term $\underset{\vphantom{l} m}{\vphantom{f} a} p_{\alpha}$ can be eliminated and for higher-order coefficients ($m > l$) the parameters in Eq.~(\ref{eq:AXCoefGeneralSolution}) change. 
The gauged solutions (symbolized by a prime) are composed in terms of the three remaining space-time directions~$\omega_{\alpha}$, $\bar{\omega}_{\alpha}$, and~$d_{\alpha}$, 
\begin{equation}
			\underset{\vphantom{l} m}{\varphi'}{}_{\alpha} 
	= \sqrt{\underset{\vphantom{l} m}{\vphantom{f} J}} 
			\left( 
				\textup{e}^{\textup{i} \underset{\vphantom{l} m}{\vphantom{f} \kappa}} \omega_{\alpha} 
			+ \textup{e}^{-\textup{i} \underset{\vphantom{l} m}{\vphantom{f} \kappa}} \bar{\omega}_{\alpha} 
			\right) 
		+ \underset{\vphantom{l} m}{\vphantom{f} \lambda} d_{\alpha}. 
	\label{eq:AXCoefGeneralSolutionGauged}
\end{equation}
When looking for the shock wave solutions, one has to find the expressions for the free parameters. 

\section{Jump Conditions}
\subsection{Series Expansion and General Jump Conditions}
The series expansion of the field equations~$\phi^{\beta} = 0$ is expressed with respect to the electromagnetic field~(\ref{eq:FieldX}). 
Evaluating every term of Born's field equations~(\ref{eq:BFieldEquations}) and combining it to one jump series yields 
\begin{eqnarray}
			0 
	&=& \phi_{\textup{B}}{}^{\beta} \nonumber \\ 
	&=& F^-{}^{\alpha \beta}{}_{,\alpha} 
		+ \underline{
			\frac{1}{b^2} 
			F^-{} 
			F^-{}^{\alpha \beta}{}_{,\alpha} } 
		- \underline{\underline{
			\frac{1}{2 b^2} 
			F^-{}^{\alpha \beta} 
			F^-{}_{,\alpha} }} 
		+ \nonumber \\ &&
		+ \sum_{m = l'}^{\infty} 
			\left( 
				\left( 
					\underset{\vphantom{l} m}{f}{}^{\alpha \beta}{}_{,\alpha} 
				+ \underline{
					\frac{1}{b^2} 
					F^-{} 
					\underset{\vphantom{l} m}{f}{}^{\alpha \beta}{}_{,\alpha} } 
				+ \underline{
					\frac{1}{b^2} 
					F^-{}^{\alpha \beta}{}_{,\alpha} 
					F^-{}_{\gamma \delta} 
					\underset{\vphantom{l} m}{f}{}^{\gamma \delta} } 
			+ {} \right.\right. \nonumber \\ && \left.\quad\left. {}
				- \underline{\underline{
					\frac{1}{2 b^2} 
					F^-{}^{\alpha \beta} 
					F^-{}_{\gamma \delta,\alpha} 
					\underset{\vphantom{l} m}{f}{}^{\gamma \delta} }} 
				- \underline{\underline{
					\frac{1}{2 b^2} 
					F^-{}^{\alpha \beta} 
					F^-{}_{\gamma \delta} 
					\underset{\vphantom{l} m}{f}{}^{\gamma \delta}{}_{,\alpha} }} 
				- \underline{\underline{
					\frac{1}{2 b^2} 
					F^-{}_{,\alpha} 
					\underset{\vphantom{l} m}{f}{}^{\alpha \beta} }} 
				\right) 
				\underset{\vphantom{l} m}{\vphantom{f} h} 
			+ {} \right. \nonumber \\ && \left. {}
			+ p_{\alpha} 
				\left( 
					\underset{\vphantom{l} m}{f}{}^{\alpha \beta} 
				+ \underline{
					\frac{1}{b^2} 
					F^-{} 
					\underset{\vphantom{l} m}{f}{}^{\alpha \beta} } 
				- \underline{\underline{
					\frac{1}{2 b^2} 
					F^-{}^{\alpha \beta} 
					F^-{}_{\gamma \delta} 
					\underset{\vphantom{l} m}{f}{}_{\gamma \delta} }} 
				\right) 
				\underset{\vphantom{l} m - 1}{\vphantom{f} h} 
			\right) 
		+ \nonumber \\ && {}
		+ \frac{1}{b^2} 
			\sum_{m, n = l'}^{\infty} 
			\left( 
				\left( 
					\underline{
					F^-{}_{\gamma \delta} 
					\underset{\vphantom{l} m}{f}{}^{\gamma \delta} 
					\underset{\vphantom{l} n}{f}{}^{\alpha \beta}{}_{,\alpha} } 
				+ \underline{
					\frac{1}{2} 
					F^-{}^{\alpha \beta}{}_{,\alpha} 
					\underset{\vphantom{l} m}{f}{}_{\gamma \delta} 
					\underset{\vphantom{l} n}{f}{}^{\gamma \delta} } 
				- \underline{\underline{
					\frac{1}{2} 
					F^-{}^{\alpha \beta} 
					\underset{\vphantom{l} m}{f}{}_{\gamma \delta,\alpha} 
					\underset{\vphantom{l} n}{f}{}^{\gamma \delta} }} 
			+ {} \right.\right. \nonumber \\ && \left.\quad\left. {}
				- \underline{\underline{
					\frac{1}{2} 
					F^-{}_{\gamma \delta,\alpha} 
					\underset{\vphantom{l} m}{f}{}^{\alpha \beta} 
					\underset{\vphantom{l} n}{f}{}^{\gamma \delta} }} 
				- \underline{\underline{
					\frac{1}{2} 
					F^-{}_{\gamma \delta} 
					\underset{\vphantom{l} m}{f}{}^{\alpha \beta} 
					\underset{\vphantom{l} n}{f}{}^{\gamma \delta}{}_{,\alpha} }} 
				\right) 
				\underset{\vphantom{l} m}{\vphantom{f} h} 
				\underset{\vphantom{l} n}{\vphantom{f} h} 
			+ {} \right. \nonumber \\ && \left. {}
			+ p_{\alpha} 
				\left( 
					\underline{
					F^-{}_{\gamma \delta} 
					\underset{\vphantom{l} m}{f}{}^{\gamma \delta} 
					\underset{\vphantom{l} n}{f}{}^{\alpha \beta} } 
				- \underline{\underline{
					\frac{1}{2} 
					F^-{}^{\alpha \beta} 
					\underset{\vphantom{l} m}{f}{}_{\gamma \delta} 
					\underset{\vphantom{l} n}{f}{}^{\gamma \delta} }} 
				- \underline{\underline{
					\frac{1}{2} 
					F^-{}_{\gamma \delta} 
					\underset{\vphantom{l} m}{f}{}^{\alpha \beta} 
					\underset{\vphantom{l} n}{f}{}^{\gamma \delta} }} 
				\right) 
				\underset{\vphantom{l} m}{\vphantom{f} h} 
				\underset{\vphantom{l} n - 1}{\vphantom{f} h} 
			\right) 
		+ \nonumber \\ && {}
		+ \frac{1}{2 b^2} 
			\sum_{m, n, o = l'}^{\infty} 
			\left(
				\left( 
					\underline{
					\underset{\vphantom{l} m}{f}{}_{\gamma \delta} 
					\underset{\vphantom{l} n}{f}{}^{\gamma \delta} 
					\underset{\vphantom{l} o}{f}{}^{\alpha \beta}{}_{,\alpha} } 
				- \underline{\underline{
					\underset{\vphantom{l} m}{f}{}^{\alpha \beta} 
					\underset{\vphantom{l} n}{f}{}_{\gamma \delta,\alpha} 
					\underset{\vphantom{l} m}{f}{}^{\gamma \delta} }} 
				\right) 
				\underset{\vphantom{l} m}{\vphantom{f} h} 
				\underset{\vphantom{l} n}{\vphantom{f} h} 
				\underset{\vphantom{l} o}{\vphantom{f} h} 
			+ {} \right. \nonumber \\ && \left. {}
			+ p_{\alpha} 
				\left( 
					\underline{
					\underset{\vphantom{l} m}{f}{}_{\gamma \delta} 
					\underset{\vphantom{l} n}{f}{}^{\gamma \delta} 
					\underset{\vphantom{l} o}{f}{}^{\alpha \beta} } 
				- \underline{\underline{
					\underset{\vphantom{l} m}{f}{}^{\alpha \beta} 
					\underset{\vphantom{l} n}{f}{}_{\gamma \delta} 
					\underset{\vphantom{l} o}{f}{}^{\gamma \delta} }} 
				\right) 
				\underset{\vphantom{l} m}{\vphantom{f} h} 
				\underset{\vphantom{l} n}{\vphantom{f} h} 
				\underset{\vphantom{l} o - 1}{\vphantom{f} h} 
			\right). 
	\label{eq:FieldXB} 
\end{eqnarray}
The full jump series expansion of the Born-Infeld field equations~$\phi_{\textup{BI}}{}^{\beta} = 0$ is given in Ref.~\cite{2014minz}. 
The series expansion of the field equations~$\phi^{\beta} = 0$ of any nonlinear electrodynamic theory follows the same mathematical formalism. 
However, analytic functions of the field invariants have to be developed in a Taylor series so that every Taylor term can be expanded in a jump series. 

The first-order jump conditions are calculated with the help of the jump brackets~(\ref{eq:JumpBrackets}), 
\begin{equation}
			0 
	= \left[ \frac{\partial^{l - 2} \phi^{\beta}}{{\partial{\varSigma}}^{l - 2}} \right]_{\varSigma = 0}, 
	\label{eq:JC1l2}
\end{equation}
higher-order jump conditions, with the index~$m > l$ are derived from similar equations, 
\begin{equation}
			0 
	= \left[ \frac{\partial^{m - 2} \phi^{\beta}}{{\partial{\varSigma}}^{m - 2}} \right]_{\varSigma = 0}. 
	\label{eq:JC1m2}
\end{equation}
We discuss the first-order jump conditions in the following. 

\subsection{First-Order Jump Conditions}
\label{sec:firstorderjumpconditions}
For the first-order jump conditions~(\ref{eq:JC1l2}) it is useful to define two background tensors. 
Their expressions are picked from the jump series expansion. 
The series expansion~(\ref{eq:FieldXB}) of the Born electrodynamics gives 
\begin{equation}
			\overset{\textup{a}}{B}{}^-_{\textup{B}} 
	:= 1 
		+ \frac{1}{b^2} F^-{} 
	\label{eq:BBackgroundTensora}
\end{equation}
and 
\begin{equation}
			\overset{\textup{b}}{B}{}^-_{\textup{B}}{}^{\alpha \beta \gamma \delta} 
	:= - \frac{1}{b^2} 
			F^-{}^{\alpha \beta} 
			F^-{}^{\gamma \delta}. 
	\label{eq:BBackgroundTensorb}
\end{equation}
When analyzing the series expansion of the Born-Infeld electrodynamics, one defines 
\begin{equation}
			\overset{\textup{a}}{B}{}^-_{\textup{BI}} 
	:= 1 
		+ \frac{1}{b^2} F^-{} 
		- \frac{1}{b^4} \left( G^- \right)^2 
	\label{eq:BIBackgroundTensora}
\end{equation}
and 
\begin{eqnarray}
			\overset{\textup{b}}{B}{}^-_{\textup{BI}}{}^{\alpha \beta \gamma \delta} 
	&:=& \frac{1}{b^2} 
			\left( 
			- F^-{}^{\alpha \beta} 
				F^-{}^{\gamma \delta} 
			- \overset{*}{F}{}^-{}^{\alpha \beta} 
				\overset{*}{F}{}^-{}^{\gamma \delta} 
			\right) 
		+ {} \nonumber \\ && {}
		+ \frac{1}{b^4} 
			\left( 
				G^- 
				\overset{*}{F}{}^-{}^{\alpha \beta} 
				F^-{}^{\gamma \delta} 
			+ G^- 
				F^-{}^{\alpha \beta} 
				\overset{*}{F}{}^-{}^{\gamma \delta} 
			- F^-{} 
				\overset{*}{F}{}^-{}^{\alpha \beta} 
				\overset{*}{F}{}^-{}^{\gamma \delta} 
			\right). 
	\label{eq:BIBackgroundTensorb}
\end{eqnarray}

In general, the first-order background field tensors read 
\begin{eqnarray}
			\overset{\textup{a}}{B}{}^- 
	&=& - 2 \left. \mathcal{L}_{F} \right|_{-}, 
	\nonumber
	\\
			\overset{\textup{b}}{B}{}^-{}^{\alpha \beta \gamma \delta}
	&=& - 4 \left. \mathcal{L}_{FF} \right|_{-} 
			F^-{}^{\alpha \beta} 
			F^-{}^{\gamma \delta} 
		- \left. \mathcal{L}_{GG} \right|_{-} 
			\overset{*}{F}{}^-{}^{\alpha \beta} 
			\overset{*}{F}{}^-{}^{\gamma \delta} 
		+ {} \nonumber \\ && {}
		- 2 \left. \mathcal{L}_{FG} \right|_{-} 
			\left( 
				F^-{}^{\alpha \beta} 
				\overset{*}{F}{}^-{}^{\gamma \delta} 
			+ \overset{*}{F}{}^-{}^{\alpha \beta} 
				F^-{}^{\gamma \delta} 
			\right). 
	\nonumber
\end{eqnarray}
The derivatives of the Lagrange density ($\mathcal{L}_{F} := \frac{\partial \mathcal{L}}{\partial F}$, $\mathcal{L}_{FF} := \frac{\partial^2 \mathcal{L}}{{\partial F}^2}$, $\mathcal{L}_{FG} := \frac{\partial^2 \mathcal{L}}{\partial F \partial G}$, and $\mathcal{L}_{GG} := \frac{\partial^2 \mathcal{L}}{{\partial G}^2}$) are evaluated with the background fields. 

The field coefficients~$\underset{l - 1}{f}{}_{\alpha \beta}$ in the first-order jump conditions are replaced according to Eq.~(\ref{eq:FieldXCoef1}) which yields 
\begin{equation}
			0 
	= \overset{\textup{a}}{B}{}^- 
			p_{\alpha} 
			\left( 
				p^{\alpha} \underset{l}{\varphi'}{}^{\beta} 
			- p^{\beta} \underset{l}{\varphi'}{}^{\alpha} 
			\right) 
		+ \overset{\textup{b}}{B}{}^-{}^{\beta \delta} 
			\underset{l}{\varphi'}{}_{\delta} 
	\label{eq:JC1l2BackgroundTensors}
\end{equation}
with the symmetric contraction 
\begin{equation}
			\overset{\textup{b}}{B}{}^-{}^{\beta \delta} 
	:= p_{\alpha} p_{\gamma} 
			\overset{\textup{b}}{B}{}^-{}^{\alpha \beta \gamma \delta}. 
	\label{eq:BackgroundTensorbContracted}
\end{equation}
The four first-order jump conditions for any electrodynamic theory derived from a Lagrange density $\mathcal{L}(F, G)$ always have the form~(\ref{eq:JC1l2BackgroundTensors}). 

The background tensor~$\overset{\textup{b}}{B}{}^-{}^{\alpha \beta \gamma \delta}$ is symmetric when exchanging the first and last index pair, 
\begin{equation}
			\overset{\textup{b}}{B}{}^-{}^{\alpha \beta \gamma \delta} 
	= \overset{\textup{b}}{B}{}^-{}^{\gamma \delta \alpha \beta}. 
	\label{eq:BackgroundTensorbSymmetries}
\end{equation}
Due to the antisymmetry of indices~$\alpha \beta$ and~$\gamma \delta$, respectively, the contracted background field tensor~$\overset{\textup{b}}{B}{}^-{}^{\alpha \beta}$ is orthogonal to the normal direction, 
\begin{equation}
			p_{\alpha} \overset{\textup{b}}{B}{}^-{}^{\alpha \beta} 
	= p_{\beta} \overset{\textup{b}}{B}{}^-{}^{\alpha \beta} 
	= 0. 
	\label{eq:BackgroundTensorbContractedProperties}
\end{equation}

Furthermore, the first-order jump conditions~(\ref{eq:JC1l2}) are a homogeneous, linear system of equations 
\begin{equation}
			N^{\alpha \beta} 
			\underset{l}{\varphi'}{}_{\beta} 
	= 0 
	\label{eq:JC1l2LSE}
\end{equation}
with the coefficient matrix 
\begin{equation}
			N^{\alpha \beta} 
	:= \overset{\textup{a}}{B}{}^- 
			\left( 
				p \eta^{\alpha \beta} 
			- p^{\alpha} p^{\beta} 
			\right) 
		+ \overset{\textup{b}}{B}{}^-{}^{\alpha \beta}. 
	\label{eq:JC1l2Matrix}
\end{equation}
Calculations of the matrix rank of~$N^{\alpha \beta}$ give the degrees of freedom of the solutions~$\underset{l}{\varphi'}{}_{\beta}$ under certain conditions yet to be identified. 
The temporal parameter~$\underset{l}{\vphantom{f} \lambda}$, the parameter~$p$, and the shock wave polarization angle~$\underset{l}{\vphantom{f} \kappa}$ are derived by contraction of Eq.~(\ref{eq:JC1l2}) with the four tetrad fields~$p_{\alpha}$, $d_{\alpha}$, $\omega_{\alpha}$, and~$\bar{\omega}_{\alpha}$. 

\subsection{First-Order Tetrad Contractions}
\label{sec:firstordertetradcontractions}
Because the first-order jump conditions~(\ref{eq:JC1l2BackgroundTensors}) do not differ for any electrodynamics when formulated in the form~(\ref{eq:JC1l2BackgroundTensors}), this section presents the general contractions before the Born and the Born-Infeld electrodynamics are investigated in detail. 

Similar to linear electrodynamics~\cite{1962treder}, the~$p_{\beta}$ contraction of the first-order jump conditions vanishes identically with the help of Eq.~(\ref{eq:BackgroundTensorbContractedProperties}), 
\begin{equation}
			p_{\beta} 
			\left[ \frac{\partial^{l - 2} \phi^{\beta}}{{\partial{\varSigma}}^{l - 2}} \right]_{\varSigma = 0} 
	\equiv 0. 
	\label{eq:JC1l2nContracted}
\end{equation}

After contracting with the timelike vector field~$d_{\beta}$, 
\begin{equation}
			0 
	= d_{\beta} 
			\left[ \frac{\partial^{l - 2} \phi^{\beta}}{{\partial{\varSigma}}^{l - 2}} \right]_{\varSigma = 0} 
	\label{eq:JC1l2tContracting}
\end{equation}
and replacing the coefficients by the gauged general solution~(\ref{eq:AXCoefGeneralSolutionGauged}), one obtains 
\begin{equation}
			0 
	= - \left( 
				\overset{\textup{a}}{B}{}^- 
				\left( d^2 - p \right) 
			- d_{\beta} d_{\delta} 
				\overset{\textup{b}}{B}{}^-{}^{\beta \delta} 
			\right) \underset{l}{\vphantom{f} \lambda} 
		+ \sqrt{\underset{l}{\vphantom{f} J}} 
			\left( 
				\textup{e}^{\textup{i} \underset{l}{\vphantom{f} \kappa}} 
				d_{\beta} \omega_{\delta} 
			+ \textup{e}^{-\textup{i} \underset{l}{\vphantom{f} \kappa}} 
				d_{\beta} \bar{\omega}_{\delta} 
			\right) 
			\overset{\textup{b}}{B}{}^-{}^{\beta \delta}. \qquad 
	\label{eq:BJC1l2tContracted}
\end{equation}
The temporal parameter has to be 
\begin{equation}
			\underset{l}{\vphantom{f} \lambda} 
	= \frac{ 
				\sqrt{\underset{l}{\vphantom{f} J}} 
				\left( 
					\textup{e}^{\textup{i} \underset{l}{\vphantom{f} \kappa}} 
					d_{\alpha} \omega_{\beta} 
				+ \textup{e}^{-\textup{i} \underset{l}{\vphantom{f} \kappa}} 
					d_{\alpha} \bar{\omega}_{\beta} 
				\right) 
				\overset{\textup{b}}{B}{}^-{}^{\alpha \beta} 
			}{
				\overset{\textup{a}}{B}{}^- 
				\left( d^2 - p \right) 
			- d_{\gamma} d_{\delta} 
				\overset{\textup{b}}{B}{}^-{}^{\gamma \delta} 
			}. 
	\label{eq:Coef1l2tcoef}
\end{equation}
It depends on the shock excitation~$\underset{l}{\vphantom{f} J}$ and the polarization angle~$\underset{l}{\vphantom{f} \kappa}$. 
In absence of background fields, $F^-{}^{\alpha \beta} = 0$, or in the linear limit~$b \to \infty$ the first jump conditions are identical to Maxwell's theory~\cite{1962treder}, with~$\underset{l}{\vphantom{f} \lambda} = 0$. 

The general condition for the parameter~$p$ follows from the contraction with the spacelike vector fields~$\omega_{\beta}$. 
The complex result splits into one real and one imaginary condition which may also constrain the polarization angle~$\underset{l}{\vphantom{f} \kappa}$ for nonlinear theories. 
The first-order parameters~$\underset{l}{\vphantom{f} J}$ and~$\underset{l}{\vphantom{f} \kappa}$ fulfill differential equations which follow from the second-order jump relations. 
However, this work focuses on the physical results of the first order. 

The~$\omega_{\alpha}$ contraction can be split into the real part 
\begin{subequations}
	\label{eq:JC1l2sReImContracted}
	\begin{eqnarray}
				0 
		&=& \overset{\textup{a-b}}{B}{}^-_d 
				\sqrt{\underset{l}{\vphantom{f} J}} 
				\left( 
					\textup{e}^{\textup{i} \underset{l}{\vphantom{f} \kappa}} \omega_{\beta} 
				+ \textup{e}^{-\textup{i} \underset{l}{\vphantom{f} \kappa}} \bar{\omega}_{\beta} 
				\right) 
				\left[ \frac{\partial^{l - 2} \phi^{\beta}}{{\partial{\varSigma}}^{l - 2}} \right]_{\varSigma = 0} 
		\nonumber
		\\
		&=& - \overset{\textup{a}}{B}{}^- 
				\overset{\textup{a-b}}{B}{}^-_d 
				p 
				\underset{l}{\vphantom{f} J} 
			+ \left( 
					\overset{\textup{a-b}}{B}{}^-_d 
					\overset{\textup{b}}{B}{}^-_\omega 
				+ \overset{\textup{bb}}{B}{}^-_{\omega d} 
				\right) 
				\underset{l}{\vphantom{f} J} 
			+ \nonumber \\ && {}
			+ \left( 
					\overset{\textup{a-b}}{B}{}^-_d 
					\overset{\textup{b}}{B}{}^-_{R\omega} 
				+ \overset{\textup{bb}}{B}{}^-_{R\omega d} 
				\right) 
				\underset{l}{\vphantom{f} J} 
				\cos{2 \underset{l}{\vphantom{f} \kappa}} 
			+ \left( 
					\overset{\textup{a-b}}{B}{}^-_d 
					\overset{\textup{b}}{B}{}^-_{I\omega} 
				+ \overset{\textup{bb}}{B}{}^-_{I\omega d} 
				\right) 
				\underset{l}{\vphantom{f} J}
				\sin{2 \underset{l}{\vphantom{f} \kappa}} 
		\label{eq:JC1l2sReContracted}
	\end{eqnarray}
	and the imaginary part 
	\begin{eqnarray}
				0 
		&=& \overset{\textup{a-b}}{B}{}^-_d 
				\sqrt{\underset{l}{\vphantom{f} J}} 
				\left( 
					\textup{e}^{\textup{i} \underset{l}{\vphantom{f} \kappa}} \omega_{\beta} 
				- \textup{e}^{-\textup{i} \underset{l}{\vphantom{f} \kappa}} \bar{\omega}_{\beta} 
				\right) 
				\left[ \frac{\partial^{l - 2} \phi^{\beta}}{{\partial{\varSigma}}^{l - 2}} \right]_{\varSigma = 0} 
		\nonumber
		\\
		&=& \textup{i} \left( 
					\overset{\textup{a-b}}{B}{}^-_d 
					\overset{\textup{b}}{B}{}^-_{R\omega} 
				+ \overset{\textup{bb}}{B}{}^-_{R\omega d} 
				\right) 
				\underset{l}{\vphantom{f} J} 
				\sin{2 \underset{l}{\vphantom{f} \kappa}} 
			- \textup{i} \left( 
					\overset{\textup{a-b}}{B}{}^-_d 
					\overset{\textup{b}}{B}{}^-_{I\omega} 
				+ \overset{\textup{bb}}{B}{}^-_{I\omega d} 
				\right) 
				\underset{l}{\vphantom{f} J} 
				\cos{2 \underset{l}{\vphantom{f} \kappa}}. 
		\label{eq:JC1l2sImContracted}
	\end{eqnarray}
\end{subequations}
The symbols used in these equations are 
{\allowdisplaybreaks
\begin{subequations}
	\label{eq:BackgroundTensorScalars}
	\begin{eqnarray}
				\overset{\textup{b}}{B}{}^-_{R\omega} 
		&:=& \left( 
					\omega_{\alpha} \omega_{\beta} 
				+ \bar{\omega}_{\alpha} \bar{\omega}_{\beta} 
				\right) 
				\overset{\textup{b}}{B}{}^-{}^{\alpha \beta}, \qquad
		\label{eq:BackgroundTensorbsReScalars}
		\\
				\overset{\textup{b}}{B}{}^-_{I\omega} 
		&:=& \textup{i} \left( 
					\omega_{\alpha} \omega_{\beta} 
				- \bar{\omega}_{\alpha} \bar{\omega}_{\beta} 
				\right) 
				\overset{\textup{b}}{B}{}^-{}^{\alpha \beta}, \qquad
		\label{eq:BackgroundTensorbsImScalars}
	\\
				\overset{\textup{b}}{B}{}^-_\omega 
		&:=& 2 \omega_{\alpha} \bar{\omega}_{\beta} 
				\overset{\textup{b}}{B}{}^-{}^{\alpha \beta}, \qquad
		\label{eq:BackgroundTensorbsScalar}
		\\
				\overset{\textup{b}}{B}{}^-_d 
		&:=& d_{\alpha} d_{\beta} 
				\overset{\textup{b}}{B}{}^-{}^{\alpha \beta}, \qquad
		\label{eq:BackgroundTensorbtScalar}
	\\
				\overset{\textup{bb}}{B}{}^-_{R\omega d} 
		&:=& \left( 
					\omega_{\alpha} \omega_{\gamma} 
				+ \bar{\omega}_{\alpha} \bar{\omega}_{\gamma} 
				\right) 
				d_{\beta} d_{\delta} 
				\overset{\textup{b}}{B}{}^-{}^{\alpha \beta} 
				\overset{\textup{b}}{B}{}^-{}^{\gamma \delta}, \qquad
		\label{eq:BackgroundTensorbbstReScalars}
		\\
				\overset{\textup{bb}}{B}{}^-_{I\omega d} 
		&:=& \textup{i} \left( 
					\omega_{\alpha} \omega_{\gamma} 
				- \bar{\omega}_{\alpha} \bar{\omega}_{\gamma} 
				\right) 
				d_{\beta} d_{\delta} 
				\overset{\textup{b}}{B}{}^-{}^{\alpha \beta} 
				\overset{\textup{b}}{B}{}^-{}^{\gamma \delta}, \qquad
		\label{eq:BackgroundTensorbbstImScalars}
	\\
				\overset{\textup{bb}}{B}{}^-_{\omega d} 
		&:=& 2 \omega_{\alpha} d_{\beta} 
				\bar{\omega}_{\gamma} d_{\delta} 
				\overset{\textup{b}}{B}{}^-{}^{\alpha \beta} 
				\overset{\textup{b}}{B}{}^-{}^{\gamma \delta}, \qquad
		\label{eq:BackgroundTensorbbstScalar}
		\\
				\overset{\textup{a-b}}{B}{}^-_d 
		&:=& \overset{\textup{a}}{B}{}^- 
				\left( d^2 - p \right) 
			- d_{\alpha} d_{\beta} 
				\overset{\textup{b}}{B}{}^-{}^{\alpha \beta}. \qquad
		\label{eq:BackgroundTensoraandbtScalar}
	\end{eqnarray}
\end{subequations}
}

If the background fields vanish, all terms in Eq.~(\ref{eq:JC1l2sReContracted}) except~$\overset{\textup{a}}{B}{}^- p \underset{l}{\vphantom{f} J}$ disappear so that the result~$p = p_{\alpha} p^{\alpha} = 0$ is identical to the linear, lightlike condition from Maxwell's electrodynamics~\cite{1962treder}. 
The Eq.~(\ref{eq:JC1l2sImContracted}) is trivially satisfied in absence of background fields. 

In general, $\underset{l}{\vphantom{f} J} \neq 0$, so that the nonlinear system of equations~(\ref{eq:JC1l2sReImContracted}) gives conditions for~$p$ and conditions for the polarization angle~$\underset{l}{\vphantom{f} \kappa}$. 

If the coefficients of the trigonometric functions in Eq.~(\ref{eq:JC1l2sImContracted}) do not vanish, it can be solved for the polarization angle. 
The nonlinear shock wave polarization condition is 
\begin{equation}
			\tan{2 \underset{l}{\vphantom{f} \kappa}} 
	= \frac{ 
				\overset{\textup{a-b}}{B}{}^-_d 
				\overset{\textup{b}}{B}{}^-_{I\omega} 
			+ \overset{\textup{bb}}{B}{}^-_{I\omega d} 
			}{ 
				\overset{\textup{a-b}}{B}{}^-_d 
				\overset{\textup{b}}{B}{}^-_{R\omega} 
			+ \overset{\textup{bb}}{B}{}^-_{R\omega d} 
			}, 
	\label{eq:Coef1PhaseConditionTan}
\end{equation}

When the trigonometric functions in Eq.~(\ref{eq:JC1l2sReContracted}) are replaced by Eq.~(\ref{eq:Coef1PhaseConditionTan}), a quadratic equation for the scalar~$p$ is obtained, 
\begin{eqnarray}
			0 
	&=& \overset{\textup{a}}{B}{}^- 
			\overset{\textup{a-b}}{B}{}^-_{d} 
			p 
		- \left( 
				\overset{\textup{a-b}}{B}{}^-_{d} 
					\overset{\textup{b}}{B}{}^-_{\omega} 
				+ \overset{\textup{bb}}{B}{}^-_{\omega d} 
			\right) 
		+ \nonumber \\ && {} 
		\pm \sqrt{ 
				\left( 
					\overset{\textup{a-b}}{B}{}^-_{d} 
					\overset{\textup{b}}{B}{}^-_{R\omega} 
				+ \overset{\textup{bb}}{B}{}^-_{R\omega d} 
				\right)^2 
			+	\left( 
					\overset{\textup{a-b}}{B}{}^-_{d} 
					\overset{\textup{b}}{B}{}^-_{I\omega} 
				+ \overset{\textup{bb}}{B}{}^-_{I\omega d} 
				\right)^2 
			}. \qquad
	\label{eq:Contraints}
\end{eqnarray}
The quadratic dependence becomes clear when replacing the background tensors~(\ref{eq:BackgroundTensorScalars}), especially $\overset{\textup{a-b}}{B}{}^-_d$ which depend on $p$ and the vector field $p_{\alpha}$ due to the contraction~(\ref{eq:BackgroundTensorbContracted}). 
The two solutions can be formulated with optical metrics, 
\begin{equation}
			g_{\textup{opt,(1,2)}}{}^{\alpha \beta} 
			p_{\alpha} p_{\beta} 
	= 0, 
	\label{eq:OpticalMetrics} 
\end{equation}
and differ in their polarization angle~(\ref{eq:Coef1PhaseConditionTan}) which also depends on the vector field~$p_{\alpha}$. 
This is called birefringence in vacuum. 
If~$g_{\textup{opt,(1)}}{}^{\alpha \beta} = g_{\textup{opt,(2)}}{}^{\alpha \beta}$, this effect does not appear. 

\subsection{Optical Metric and Shock Wave Polarization of the Born Electrodynamics}
Since the Born electrodynamics~(\ref{eq:BLagrange}) has a background tensor~$\overset{\textup{b}}{B}{}^-{}^{\alpha \beta \gamma \delta} \propto F^-{}^{\alpha \beta} F^-{}^{\gamma \delta}$, the contracted background tensor~(\ref{eq:BackgroundTensorbContracted}) shows the symmetry
\begin{equation}
			\overset{\textup{b}}{B}{}^-{}^{\alpha \beta} 
			\overset{\textup{b}}{B}{}^-{}^{\gamma \delta} 
	= \overset{\textup{b}}{B}{}^-{}^{\alpha \delta} 
			\overset{\textup{b}}{B}{}^-{}^{\gamma \beta}. 
	\label{eq:BackgroundTensorbNonlinearSymmetry}
\end{equation}
Contractions with different combinations of the timelike and spacelike tetrads yield three identities, 
\begin{eqnarray}
			\overset{\textup{b}}{B}{}^-_\omega 
			\overset{\textup{b}}{B}{}^-_d 
	&=& \overset{\textup{bb}}{B}{}^-_{\omega d}, 
	\label{eq:BackgroundTensorbNonlinearSymmetryts}
	\\
			\overset{\textup{b}}{B}{}^-_{R\omega} 
			\overset{\textup{b}}{B}{}^-_d 
	&=& \overset{\textup{bb}}{B}{}^-_{R\omega d}, 
	\label{eq:BackgroundTensorbNonlinearSymmetrytsR}
	\\
			\overset{\textup{b}}{B}{}^-_{I\omega} 
			\overset{\textup{b}}{B}{}^-_d 
	&=& \overset{\textup{bb}}{B}{}^-_{I\omega d}. 
	\label{eq:BackgroundTensorbNonlinearSymmetrytsI}
\end{eqnarray}
These three equations and the phase condition~(\ref{eq:Coef1PhaseConditionTan}) yield 
\begin{equation}
			\tan{2 \underset{l}{\vphantom{f} \kappa}} 
	= \frac{ 
				\overset{\textup{b}}{B}{}^-_{I\omega} 
			}{ 
				\overset{\textup{b}}{B}{}^-_{R\omega} 
			} 
	= \left( 
			\frac{\textup{Re}(\omega_{\alpha}) p_{\beta} F^-{}^{\alpha \beta}} 
				{\textup{Im}(\omega_{\gamma}) p_{\delta} F^-{}^{\gamma \delta}} 
		\right)^2. 
	\label{eq:BCoef1PhaseCondition}
\end{equation}
The right-hand side with the squared bracket is the result for any nonlinear theory~$\mathcal{L}(F)$, including the Born electrodynamics. 

With the above symmetry, the quadratic equation~(\ref{eq:Contraints}) becomes 
\begin{equation}
			\overset{\textup{a}}{B}{}^- 
			p 
		- \frac{p}{d^2 - p} 
			\overset{\textup{b}}{B}{}^-_d 
		- (1 \pm 1) \overset{\textup{b}}{B}{}^-_\omega 
	= 0. 
	\label{eq:BJC1l2sReContractedSimplified}
\end{equation}

The equation with the upper sign is fulfilled for the parameter~$p_{(1)} = 0$ (known from Maxwell's electrodynamics) and the first optical metric is identical to the Minkowskian metric, 
\begin{equation}
			g_{\textup{opt,(1)}}{}^{\alpha \beta} 
	= \eta^{\alpha \beta}. 
	\label{eq:OpticalMetric1}
\end{equation}

When contracting the second background tensor with the Minkowskian metric given in Eq.~(\ref{eq:TetradsMetric}), the second equation with the minus sign can be rewritten to 
\begin{equation}
			\underbrace{\left( 
				\overset{\textup{a}}{B}{}^- 
				\eta^{\alpha \beta} 
			+ \overset{\textup{b}}{B}{}^-{}^{\gamma \alpha}{}_{\gamma}{}^{\beta} 
			\right)}_{g_{\textup{opt,(2)}}{}^{\alpha \beta}} 
			p_{\alpha} p_{\beta} 
	= 0. 
	\label{eq:OpticalMetric2}
\end{equation}
The symmetric, second-order tensor in the brackets is the second optical metric. 
With the background tensors of the Born electrodynamics, one finds 
\begin{equation}
			g_{\textup{opt,B,(2)}}{}^{\alpha \beta} 
	= \left( 
				1 
			+ \frac{1}{b^2} F^-{} 
			\right) 
			\eta^{\alpha \beta} 
		+ \frac{1}{b^2} 
			F^-{}^{\alpha \gamma} 
			F^-{}_{\gamma}{}^{\beta}. 
	\label{eq:BOpticalMetric2}
\end{equation}
The symmetry~(\ref{eq:BackgroundTensorbNonlinearSymmetry}) also occurs for field theories with the Lagrange density~$\mathcal{L}(F)$ or~$\mathcal{L}(G)$. 
For $\mathcal{L}(G)$~theories, however, the background tensor~$\overset{\textup{a}}{B}{}^-$ equals zero which is nonphysical because the second optical metric would vanish in absence of background fields (for further arguments, see Ref.~\cite{2002obukhovrubilar}). 

The result~(\ref{eq:BOpticalMetric2}) was published by Novello et al.~\cite{2000novelloetal} as the only solution. 
The birefringence of the Born theory, including the first solution~(\ref{eq:OpticalMetric1}), was also found by Obukhov et al.~\cite{2002obukhovrubilar}. 

Furthermore, the shock wave polarization condition~(\ref{eq:BCoef1PhaseCondition}) has to be fulfilled for both solutions, individually. 
Two different solutions for the vector field~$p_{\alpha}{}_{(1,2)} = \varSigma_{,\alpha}{}_{(1,2)}$ are defined by the optical metrics~(\ref{eq:OpticalMetric1}) and (\ref{eq:OpticalMetric2}). 
The vector fields are used in the nominator and denominator contractions of Eq.~(\ref{eq:BCoef1PhaseCondition}) yielding two different polarizations~$\underset{l}{\vphantom{f} \kappa}{}_{(1,2)}$. 

As an example, considering a shock wave with $(p_{\alpha}) = (1, \sqrt{1 - p_{(1,2)}}, 0, 0)$, $(d_{\alpha}) = (1, 0, 0, 0)$, and $(\omega_{\alpha}) = (0, 0, 1/2, \textup{i}/2)$, the polarization condition reads (in the International System of Units) 
\begin{equation}
			\tan{2 \underset{l}{\vphantom{f} \kappa}{}_{(1,2)}} 
	= \left( 
			\frac{\frac{1}{c} \mathcal{E}_z^- + \sqrt{1 - p_{(1,2)}} \mathcal{B}_y^-} 
				{\frac{1}{c} \mathcal{E}_y^- + \sqrt{1 - p_{(1,2)}} \mathcal{B}_z^-} 
		\right)^2. 
	\label{eq:BCoef1PhaseConditionX}
\end{equation}
In an experimental setup with static, electric background fields~$\mathcal{E}_y^-$ and~$\mathcal{E}_z^-$ and static, magnetic background fields~$\mathcal{B}_y^-$ and~$\mathcal{B}_z^-$, two different polarizations should appear. 
If either both electric or both magnetic background fields are zero, both polarizations are identical. 

The polarization condition~(\ref{eq:BCoef1PhaseCondition}) is supported by the fact that the rank of the coefficient matrix~(\ref{eq:JC1l2Matrix}) is reduced to 
\begin{equation}
			\textup{rank} \left( N_{\textup{B}}{}^{\alpha \beta} \right) 
	= 2, 
	\label{eq:BJC1MatrixRank}
\end{equation}
with the help of the optical metric~(\ref{eq:OpticalMetric1}) or (\ref{eq:BOpticalMetric2}), respectively. 
Therefore, the \emph{gauged} first-order coefficient~(\ref{eq:AXCoefGeneralSolutionGauged}) has one free parameter, the shock excitation~$\underset{l}{\vphantom{f} J}$. 

\subsection{Optical Metric and Shock Wave Polarization of the Born-Infeld Electrodynamics}
Both optical metrics coincide for the Born-Infeld electrodynamics and have the same mathematical form as the second solution in the Born theory,  
\begin{equation}
			g_{\textup{opt,BI,(1,2)}}{}^{\alpha \beta} 
	= \left( 
				1 
			+ \frac{1}{b^2} F^-{} 
			\right) 
			\eta^{\alpha \beta} 
		+ \frac{1}{b^2} 
			F^-{}^{\alpha \gamma} 
			F^-{}_{\gamma}{}^{\beta}. 
	\label{eq:BIOpticalMetric}
\end{equation}
These optical metrics reduce the rank of the coefficient matrix~(\ref{eq:JC1l2Matrix}) to one, 
\begin{equation}
			\textup{rank} \left( N_{\textup{BI}}{}^{\alpha \beta} \right) 
	= 1. 
	\label{eq:BJC1MatrixRankExtended}
\end{equation}
Therefore, the \emph{gauged} first-order solution has two free parameters. 
Not only the shock excitation~$\underset{l}{\vphantom{f} J}$ but also the shock polarization angle~$\underset{l}{\vphantom{f} \kappa}$ are free parameters. 
When evaluating the general equations from above, the polarization condition~(\ref{eq:Coef1PhaseConditionTan}) becomes not applicable. 

\section{Summary}
\label{sec:summary}
We considered the first-order jump relations of the shock wave formalism~\cite{1962treder} in nonlinear electrodynamics. 
The shock coefficients of the suitable gauged electromagnetic potential~$A_{\alpha}$ are obtained with a generalized Stellmacher tetrad~\cite{1938stellmacher} which is compatible with the shock wave front~$\varSigma = 0$. 
It turns out that, in general, nonlinear theories of electrodynamics show birefringence in vacuum as it has been demonstrated with other methods~\cite{1970boillat,1972boillat,2004boillatruggeri,2002obukhovrubilar,2011perlick}, too. 
As a new result, the general condition~(\ref{eq:Coef1PhaseConditionTan}) for the first-order shock polarization of Born-type electrodynamics is calculated. 

As an example, the Born electrodynamics is investigated for which the shock wave has two polarization modes traveling along different directions (birefringence). 
One of them is identical to Maxwell's theory of light in vacuum. 
The other polarization has a shock wave front with a normal vector field~$p_{\alpha} = \partial_{\alpha} \varSigma$ characterized by Eq.~(\ref{eq:OpticalMetrics}) with the optical metric~(\ref{eq:BOpticalMetric2}). 

The Born-Infeld electrodynamics, however, is a special case which does not show the effect of birefringence. 
As in the linear theory of Maxwell~\cite{1962treder}, the shock waves with arbitrary polarizations share one optical metric. 
One significant difference between the Born-Infeld and Maxwell's electrodynamics is caused by different optical metrics. 
The optical metric of the Born-Infeld electrodynamics is given by Eq.~(\ref{eq:BIOpticalMetric}). 

Our focus was on the Born and the Born-Infeld electrodynamics where especially the latter (i) allows to describe classical electrons without singularities in the electromagnetic fields, (ii) is according to Born and Infeld a concept of a unified theory for gravitational and electrodynamic fields~\cite{1934borninfeld}, and (iii) appears as effective electrodynamics in superstring theory~\cite{1987bergshoeffetal,1987metsaevetal}. 
However, calculations especially in subsections~\ref{sec:firstorderjumpconditions} and~\ref{sec:firstordertetradcontractions} are also valid for nonlinear electrodynamics of the Pleba\'{n}ski type~\cite{1970plebanski} and in particular for the Heisenberg-Euler theory where there are experimental arguments~\cite{2012dunne} in favor for it. 

\section{Acknowledgments}
\label{sec:acknowledgments}
We thank the reviewers of the journal for their comments and remarks which helped to improve this article. 

\bibliographystyle{elsarticle-num} 
\bibliography{elsmanuscriptReferences}

\begin{thebibliography}{10}
\expandafter\ifx\csname url\endcsname\relax
  \def\url#1{\texttt{#1}}\fi
\expandafter\ifx\csname urlprefix\endcsname\relax\def\urlprefix{URL }\fi
\expandafter\ifx\csname href\endcsname\relax
  \def\href#1#2{#2} \def\path#1{#1}\fi

\bibitem{1933born}
M.~Born, {Modified Field Equations with a Finite Radius of the Electron},
  Nature 132~(282) (1933) 282.
\newblock \href {http://dx.doi.org/10.1038/132282a0}
  {\path{doi:10.1038/132282a0}}.

\bibitem{1933borninfeld}
M.~Born, L.~Infeld, {Electromagnetic Mass}, Nature 132~(970) (1933) 970.
\newblock \href {http://dx.doi.org/10.1038/132970a0}
  {\path{doi:10.1038/132970a0}}.

\bibitem{1934borninfeld}
M.~Born, L.~Infeld, {Foundations of the New Field Theory}, Proceedings of the
  Royal Society of London A 144~(852) (1934) 425.
\newblock \href {http://dx.doi.org/10.1098/rspa.1934.0059}
  {\path{doi:10.1098/rspa.1934.0059}}.

\bibitem{1936heisenbergeuler}
W.~Heisenberg, H.~Euler, {Folgerungen aus der Diracschen Theorie des
  Positrons}, Zeitschrift f{\"u}r Physik 98~(11-12) (1936) 714.
\newblock \href {http://dx.doi.org/10.1007/BF01343663}
  {\path{doi:10.1007/BF01343663}}.

\bibitem{1987bergshoeffetal}
E.~Bergshoeff, E.~Sezgin, C.~N. Pope, P.~K. Townsend, {The Born-Infeld Action
  from Conformal Invariance of the Open Superstring}, Physics Letters B 188~(1)
  (1987) 70.
\newblock \href {http://dx.doi.org/10.1016/0370-2693(87)90707-6}
  {\path{doi:10.1016/0370-2693(87)90707-6}}.

\bibitem{1987metsaevetal}
R.~R. Metsaev, M.~A. Rahmanov, A.~A. Tseytlin, {The Born-Infeld Action as the
  Effective Action in the Open Superstring Theory}, Physics Letters B 193~(2-3)
  (1987) 207.
\newblock \href {http://dx.doi.org/10.1016/0370-2693(87)91223-8}
  {\path{doi:10.1016/0370-2693(87)91223-8}}.

\bibitem{1970boillat}
G.~Boillat, {Nonlinear Electrodynamics: Lagrangians and Equations of Motion},
  Journal of Mathematical Physics 11 (1970) 941.
\newblock \href {http://dx.doi.org/10.1063/1.1665231}
  {\path{doi:10.1063/1.1665231}}.

\bibitem{1972boillat}
G.~Boillat, {Shock Relations in Nonlinear Electrodynamics}, Physics Letters A
  40~(1) (1972) 9.
\newblock \href {http://dx.doi.org/10.1016/0375-9601(72)90174-0}
  {\path{doi:10.1016/0375-9601(72)90174-0}}.

\bibitem{2004boillatruggeri}
G.~Boillat, T.~Ruggeri, {Energy Momentum, Wave Velocities and Characteristic
  Shocks in Euler's Variational Equations with Application to the Born-Infeld
  Theory}, Journal of Mathematical Physics 45~(2004) (2004) 3468.
\newblock \href {http://dx.doi.org/10.1063/1.1780611}
  {\path{doi:10.1063/1.1780611}}.

\bibitem{2003hehlobukhov}
F.~W. Hehl, Y.~N. Obukhov, {Foundations of Classical Electrodynamics: Charge,
  Flux, and Metric}, Birkh\"{a}user, Boston, 2003.

\bibitem{2011perlick}
V.~Perlick, {On the Hyperbolicity of Maxwell's Equations with a Local
  Constitutive Law}, Journal of Mathematical Physics 52~(4) (2011) 042903.
\newblock \href {http://dx.doi.org/10.1063/1.3579133}
  {\path{doi:10.1063/1.3579133}}.

\bibitem{1903hadamard}
J.~Hadamard, {Le\c{c}ons sur la Propagation des Ondes et les \'{E}quations de
  L'Hydrodynamique}, A. Hermann, Paris, 1903, [translated by D. Delphenich, J.
  Hadamard, Propagation of Waves and the Equations of Hydrodynamics,
  Birkh\"{a}user, Basel, 2011].

\bibitem{1937couranthilbert}
R.~Courant, D.~Hilbert, Methoden der mathematischen Physik, Vol.~2, Julius
  Springer, Berlin, 1937.

\bibitem{1962couranthilbert}
R.~Courant, D.~Hilbert, Methods of Mathematical Physics, Vol. 2, Partial
  Differential Equations, Interscience Publishers, New York, 1962.

\bibitem{1960truesdelltoupin}
C.~Truesdell, R.~Toupin, {Encyclopedia of Physics: The Classical Field Theories
  - Singular Surfaces and Waves}, Vol. III/1, Springer, Heidelberg, 1960.

\bibitem{1962treder}
H.-J. Treder, {Gravitative Sto{\ss}wellen - Nichtanalytische Wellenl{\"o}sungen
  der Einsteinschen Gravitationsgleichungen}, Akademie-Verlag Berlin, 1962.

\bibitem{1938stellmacher}
K.~Stellmacher, {Ausbreitungsgesetze f{\"u}r charakteristische
  Singularit{\"a}ten der Gravitationsgleichungen}, Mathematische Annalen
  115~(1) (1938) 740.
\newblock \href {http://dx.doi.org/10.1007/bf01448968}
  {\path{doi:10.1007/bf01448968}}.

\bibitem{2014minz}
C.~Minz, {Shock Waves in Nonlinear Electrodynamics}, Master's thesis,
  Technische Universit\"{a}t Berlin (March 2014).

\bibitem{2002obukhovrubilar}
Y.~N. Obukhov, G.~F. Rubilar, {Fresnel Analysis of Wave Propagation in
  Nonlinear Electrodynamics}, Physics Review D 66 (2002) 024042.
\newblock \href {http://dx.doi.org/10.1103/PhysRevD.66.024042}
  {\path{doi:10.1103/PhysRevD.66.024042}}.

\bibitem{2000novelloetal}
M.~Novello, V.~A. De~Lorenci, J.~M. Salim, R.~Klippert, {Geometrical Aspects of
  Light Propagation in Nonlinear Electrodynamics}, Physical Review D 61 (2000)
  045001.
\newblock \href {http://dx.doi.org/10.1103/PhysRevD.61.045001}
  {\path{doi:10.1103/PhysRevD.61.045001}}.

\bibitem{1999jackson}
J.~D. Jackson, {Classical Electrodynamics}, 3rd Edition, John Wiley \& Sons,
  New York, 1999.

\bibitem{1961papapetroutreder}
A.~Papapetrou, H.~Treder, {Das Sprungproblem nullter Ordnung in der Allgemeinen
  Relativit{\"a}tstheorie}, Mathematische Nachrichten 23 (1961) 371.

\bibitem{1970plebanski}
J.~Pleba\'{n}ski, {Lectures on Non-Linear Electrodynamics}, Nordita,
  Copenhagen, 1970.

\bibitem{2012dunne}
G.~V. Dunne, {The Heisenberg-Euler Effective Action: 75 Years On},
  International Journal of Modern Physics: Conference Series 14~(2) (2012) 42.
\newblock \href {http://dx.doi.org/10.1142/S2010194512007222}
  {\path{doi:10.1142/S2010194512007222}}.

\end{thebibliography}

\end{document}